# Notebook-as-a-VRE (NaaVRE): from private notebooks to a collaborative cloud virtual research environment

Zhiming Zhao[1,2], Spiros Koulouzis[1,2], Riccardo Bianchi[1,2], Siamak Farshidi[1], Zeshun Shi[1], Ruyue Xin[1], Yuandou Wang[1], Na Li[1], Yifang Shi[2,3], Joris Timmermans[2,3], W. Daniel Kissling[2,3].

[1]Multiscale Networked Systems, University of Amsterdam
[2]LifeWatch ERIC, Virtual Lab & Innovation Center (VLIC)
[3]Institute for Biodiversity and Ecosystem Dynamics (IBED), University of Amsterdam
Science Park 904, 1098XH, Amsterdam, the Netherlands
{z.zhao | s.koulouzis | r. Bianchi | s.farshidi | z.shi2 | r.xin | y.wang8 | n.li | y.shi | j.timmermans | w.d.kissling}@uva.nl

*Abstract*— Virtual Research Environments (VREs) provide user-centric support in the lifecycle of research activities, e.g., discovering and accessing research assets, or composing and executing application workflows. A typical VRE is often implemented as an integrated environment, including a catalog of research assets, a workflow management system, a data management framework, and tools for enabling collaboration among users. In contrast, notebook environments like Jupyter allow researchers to rapidly prototype scientific code and share their experiments as online accessible notebooks. Jupyter can support several popular languages used by data scientists, such as Python, R, and Julia. However, such notebook environments do not have seamless support for running heavy computations on remote infrastructure or finding and accessing collaborative software code inside notebooks. This paper investigates the gap between a notebook environment and a VRE and proposes an embedded VRE solution for the Jupyter environment called Notebook-as-a-VRE (NaaVRE). The NaaVRE solution provides functional components via a component marketplace and allows users to create a customized VRE on top of the Jupyter environment. From the VRE, a user can search research assets (data, software, and algorithms), compose workflows, manage the lifecycle of an experiment, and share the results among users in the community. We demonstrate how such a solution can enhance a legacy workflow that uses Light Detection and Ranging (LiDAR) data from country-wide airborne laser scanning surveys for deriving geospatial data products of ecosystem structure at high resolution over broad spatial extents. This enables users to scale out the processing of multi-terabyte LiDAR point clouds for ecological applications to more data sources in a distributed cloud environment.

## I. Introduction

The study of many scientific problems, e.g., big environmental challenges or cancer diagnosis, requires large data volumes, advanced modeling techniques, and distributed computing facilities [1,2]. To conduct such investigations, researchers often have to reuse research assets, e.g., observational data or medical images, AI models, workflows, and infrastructure services from different parties for building computational experiments. Specifically, researchers need effective collaborative *support environments* for conducting advanced data sciences research [3]: discovery, access, interoperation, and re-use of the research assets, and integration of all resources into cohesive observational, experimental, and simulation investigations with replicable workflows.

Virtual Research Environments (VREs) support scientists in the life cycle of their research activities by enabling effective discovery and selection of data, software services, and other relevant research assets from different sources, construction of cohesive workflows, and collaboration with other scientists [4]. VREs are also called Virtual Laboratories or Science Gateways [5, 22]. Examples include VRE4EIC[1], D4Science[2], EVER-EST[3], and Galaxy[4]. Graphical environments, workflow management systems, and data analytics tools are typical components of such environments. The development of a VRE is often driven by practices from a specific user community, e.g., for managing scientific workflows and sharing their research results, resulting in specific graphical environments, workflow management systems, and data analytics tools to be components of the VRE. Furthermore, a VRE is often implemented as an integrated environment that provides users with pre-configured sources of data and software tools, and functional components for managing research activities. While providing many benefits to the scientific community, the adoption of such integrated VREs is often hampered by the high time investment for learning the new technologies and incompatible experiences, e.g., managing scientific workflows.

A notebook environment, such as Jupyter[5] (in contrast to a VRE), allows researchers to more quickly and effectively initiate their research activities, such as implement the experimental logic using scripting languages (such as Python, R, and Julia) to document and share the experiments with necessary inputs/outputs and parameters as self-contained documents (namely notebooks) [6]. Compared to integrated

---

[1] https://vre4eic.ercim.eu/
[2] https://www.d4science.org
[3] https://ever-est.eu/
[4] https://usegalaxy.org/
[5] https://jupyter.org/

VREs that require online access control (e.g., D4Science), or heavy client-side software on the local machine (e.g., for workflow management), notebook environments such as Jupyter, demonstrate their advantages of light, portable, and easy to use. As such, it remains presently the most often used method for performing research, even though the approach faces challenges of utilizing remote infrastructure [7] and restricts collaborative research designs inherent to a VRE.

There is a clear need for improving the Jupyter environment to meet the requirements for open sciences. In this paper, we first review the current VRE support of the Jupyter environment and identify the gaps. After that, we present a lightweight solution to extend a Jupyter notebook environment as a collaborative Virtual Research Environment, which provides features for effectively managing research resources, remote cloud automation, and workflows. Finally, we demonstrate the key features of our solution, using a use case that bridges ecology and remote sensing, i.e., processing multi-terabyte point clouds from national airborne laser scanning surveys to derive high-resolution geospatial layers of ecosystem structure over broad spatial extents.

II. PROBLEM DESCRIPTION AND RELATED WORK

Research activities in data-centric sciences are often a combination of many steps. We can often identify several closely related activities centered around domain-specific *scientific research*, *data management*, and *infrastructure utilization* for remote computation (Fig. 1).

Figure 1. Different research activities that might be involved in data-intensive research[6].

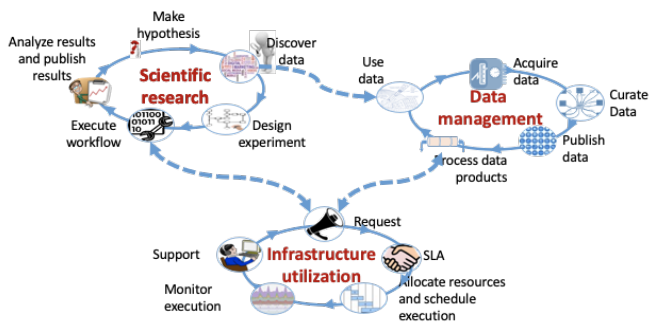

For instance,
1. When doing **scientific research** on a specific problem, a scientist often starts with *making a hypothesis* for a scientific problem and then *discovers and acquires data and software components* that can be used to *design an experiment or analysis* for the study. After that, the scientist would process the data by *executing a workflow* with specific input data and configuration of parameters of the workflow steps (e.g., via algorithms or tools). When the data set is large, such an experiment or analysis often requires a large computing and storage capacity that remote computing facilities can only provide. After the execution, the scientist then analyzes and publishes the results. In general, this traditional scientific research is not a simple sequential flow, but follows an iterative approach; for instance, a scientist can return to their original workflow design, execution, and analysis steps multiple times. Depending on specific practices, those activities can be structured in different orders or combinations, but the main components of this research life cycle remain the same [18].
2. **Data management** involves a set of activities that make data a valuable resource for subsequent (re-)use. These activities are often conducted by dedicated data managers or support staff. For instance, they start by *acquiring data* from remotely deployed sensors, in-situ field measurements, observation networks, or other data infrastructures. Experts will *curate those data* via steps like quality control, metadata annotation, and archiving. Those curated data can then be *published* and made accessible to external users via metadata catalogs or data repositories. In this context, data can also be processed for different products (*process data products*) to serve dedicated users, e.g., early warning platforms often need near-real-time data products from environmental observation networks to be quickly accessible in their big simulation platforms. Finally, there are services to make such data (either raw data or derived data products) adequate for new scientific experiments or analyses (*use data*). The activities in data management thus often differ among domains but can be generalized in typical steps of a data life cycle [17]. Those five basic activities illustrated here (Fig. 1) are common ones we observed from research infrastructures in the environmental and earth science domains and modeled in the ENVRI Reference Model[7].
3. **Infrastructure utilization-related** activities can be typically seen in data centers or computing centers for scientific research. The users who want to run scientific workflows on such infrastructures first have to *request* a certain capacity of resources from the infrastructure based on the requirement of the workflow (e.g., type of computation and volume of the data). If the requests can be handled, the infrastructure operator typically makes an agreement, also called *Service Level Agreement (SLA)* with the user. After that, the infrastructure operator will *allocate the resources* and *schedule the execution* of the request on the infrastructure. During the execution, the operator will also actively *monitor* the status of the execution and provide *support* to the users when specific errors occur (e.g., failure of the workflow).

The boundaries among those cycles in many cases are often blurred. For instance, a domain researcher often spends lots of time on data acquisition, quality control, and processing during their daily activities, as we can observe from ecologists. In the ecology domain, data acquisition and curation are also seen as required steps in scientific research. Moreover, some of the activities can be outsourced to third parties, or largely automated, e.g., infrastructure utilization, or generating specific data products. Additionally, some activities are still very difficult to be supported by a single IT system, because of their high dependencies on human actions. Nevertheless, we can clearly see the complexity a scientist faces when managing IT-related activities in the research lifecycle; an effective research support system is needed.

---

[6] The lifecycle is derived from typical activities in workflow management, data management (based on ENVRI reference model), and cloud resource management. There are certainly other ways to describe them.
[7] https://envri.eu/envri-rm-v2-0-released-today/

*A. Virtual research environment and research lifecycle*

To effectively support scientists to conduct those research activities, a VRE should provide tools and platforms to:

*Manage research assets* (data, software, and other types of objects): e.g., hide the complexity of discovering research assets across different sources, provide interfaces to the other relevant infrastructures (e.g., research infrastructures in the ENVRI cluster[8]), and curate, publish and share the newly created research assets.

*Manage scientific experiments*: compose the logic (workflows) of the scientific experiments using available components, execute the workflow locally or using remote infrastructure, and analyze the final results.

*Manage collaboration with users from the community*, e.g., for sharing research assets and cooperative work on similar problems.

To enable those basic functionalities, a VRE needs to

1. *Be flexible and efficient*: the VRE should first allow users to flexibly conduct their research activities, preferably without changing much of their existing daily practices. It means the tools and functionality provided by the VRE should be customized to the specific domain and problems that researchers are working on.
2. *Automate the standardized processes* or the processes that scientists clearly understand but do not wish to iterate much time on, e.g., data clearing pipelines and preprocessing tasks.
3. *Be reproducible* of the processes is crucial for the scientists to analyze the experiment processes, in particular when an experiment is distributed on a large scale.
4. *Provide trustworthiness on* the quality and content of data products when they are from an external source and on the services and model used for processing the data.

*B. Limits of the Jupyter environment*

As an essential type of research asset, notebooks allow users to effectively reproduce the experiments and to develop them for new purposes further. Still, there are several challenges to being an effective VRE for enabling open sciences and innovations across disciplines on a large scale:

1. *Difficult to find and access* a notebook at the granularity level of functions and components. Researchers often share their Notebooks via version control systems, like Git; insufficient metadata of the notebook, particularly the monolithic nature of the notebooks, hampers the discovery of the latter, especially the components (namely Cells) in the notebook.

2. *Lack of flexibility* to reuse a notebook as part of a distributed workflow. When reusing an existing notebook, a researcher often needs part of the code fragments and integrates it as part of the other code to build logic for a new experiment. The often tightly coupled functions in the notebook, in particular, the implicit dependencies on the libraries make the reusability and extension of the code fragments difficult.

3. *Difficult to scale* the notebook to remote infrastructures. Current notebook environments, like Jupyter Hub[9], couple a pre-configured infrastructure (e.g., via Cloud IaaS) to dynamically load instances of a notebook and perform computing tasks [8]. When processing huge data volumes or computing complex tasks, dynamically allocated cloud resources other than that pre-configured capacity are often needed for parallelizing distributed computing tasks. Solutions to seamlessly automate the infrastructure services with the scheduling of tasks in the notebook are still lacking or not widely adopted.

*C. VRE development challenges*

As a special kind of software, the commercial market of VREs is still quite small; most of the developments still originate from the academic and research communities. This also brings lots of challenges to the sustainability of the VREs.

First, the development of the VRE interface is time-consuming and often customized to research assets in a specific domain, or a specific type of computing infrastructure. It often creates gaps between the VRE and the daily practices of a scientific domain. Second, the openness of the VRE is crucial for a scientist; not only a connection to the new infrastructure is needed for scaling experiments out, but also the connection to new communities for expanding collaboration to different scientific domains. Moreover, the interface between VRE and the underlying research infrastructures (for accessing data, services, and other research assets) and computing infrastructure (for performing computing tasks) requires standardized metadata and APIs offered by those infrastructures. However, the low interoperability among current research infrastructures, makes such interfaces non-trivial.

To tackle those challenges, we propose a VRE solution that can be embedded in the daily practices which current scientists perform in their research activity and aim to expand their daily environment to a VRE with minimal effort. An embedded VRE solution called Notebook-as-a-VRE (NaaVRE) is proposed.

*D. Related Work*

There are several Jupyter-based systems [19-21] aimed to improve the notebook environment for supporting scientific experiments. The Earth Observation Data Access Gateway (EODAG) [19] Jupyter extension[10] enables end-users to query satellite imagery from various image providers, using the EODAG library via notebook environment. The results of the query can be transformed to code cells into the active Python notebook to further process the dataset. This extension focuses only on integrating a specific data resource with Jupyter. Binder [20] can automate the launch of a Jupyter notebook from a Git repository and thus simplify the reproduction of the Jupyter notebooks-based data experiments. As a highly interactive environment, cells in a notebook are often developed, tested, and executed in an arbitrary order by the user, which may result in less reusability and reproducibility of the notebook. Wang et al. [21] used code analysis techniques to analyze the dependencies among cells and to restore the reproducibility of the notebook. These existing examples address specific aspects of the scientific experiments, e.g., data access, query, automation of the execution, or reproducibility of cells based on dependencies. These solutions do not aim to cover the entire lifecycle of an experiment from a Virtual Research Environment point of view.

---

[8] www.envri.eu
[9] https://jupyter.org/hub
[10] https://github.com/CS-SI/eodag-labextension

### III. NOTEBOOK-AS-A-VRE (NAAVRE)

The Notebook-as-a-VRE (NaaVRE) solution focuses on building a VRE by extending the notebook environment of the end-users. Currently, we design it based on Jupyter [6], a widely used tool for daily research activities in lots of data-centric sciences. In principle, other environments like R-Studio can also be considered. In Jupyter, popular programming languages used by data scientists including Python, R, Julia can be supported by specific execution kernels.

*A. Requirements*

The proposed solution should meet the following requirements:
1. Should be integrated into the Jupyter environment, without forcing users to change their basic practices on conducting experiments.
2. Seamless support for adding VRE features, e.g., publishing/sharing components with the community, running tasks on remote cloud infrastructure, and search tools for new data sources.
3. Open for new functional components. The NaaVRE should further be exploited as an open ecosystem to include new community contributed components.
4. The solution should be highly customizable and allow a user to reconfigure the environment based on the needs.
5. The solution should be efficient and fault-tolerant; users in the NaaVRE ecosystem are connected via a decentralized framework for sharing data and other research assets.

*B. System architecture*

Based on the requirements, we design the NaaVRE as an ecosystem, which provides required VRE functional components as scalable services (e.g., microservices), and can be deployed as Jupyter extensions on the user environment (Fig. 2).

Figure 2. NaaVRE and the Jupyter client.

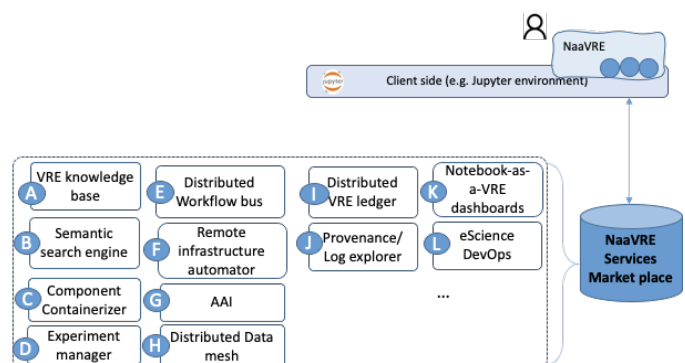

The NaaVRE provides the functional components (services) via a repository (also called marketplace) which allows users to browse and select functional components to make a customized VRE experience.

Typical functional components include:
A. **VRE knowledge base** with published research assets (data sets, software components, and tools) produced and shared by the VRE community.
B. **The semantic search engine** provides a web interface to do a search from external catalogs (e.g., for data sets, services, or other assets), and also API for the application to invoke via Jupyter.
C. **Component containerizer** as a tool enabling users to effectively select code fragments (cells) from the notebook, generate reusable workflow building blocks (e.g., RESTful services), and encapsulate the building blocks as deployable containers (e.g., Docker).
D. **Experiment manager** allows users to design experiments by selecting building blocks from the VRE knowledge base, composing the logical relations and dependencies among components, configuring the parameters, input, and output, executing the experiment via workflow bus, and monitoring the progress of the experiment.
E. **Distributed workflow bus** uses a software bus concept [15] to manage runtime infrastructures for automating the deployment of workflow instances (namely experiments), scheduling the execution of each experiment, and managing information of different workflow instances over *the distributed ledger* and *the distributed data mesh*. The distributed workflow bus enables collaboration among workflows from different users, e.g., publishing/subscribing experiment results, and reproducing a specific result.
F. **Remote Infrastructure Automator** that automates the execution of a workflow on the remote infrastructure, including a) plan the cloud infrastructure (networked virtual machines) based on workflow description, providers, and execution requirements, b) automates the provision of virtual machines, the configuration of Kubernetes clusters, and c) schedules the execution of the workflow.
G. **Authentication and authorization infrastructure (AAI)** [9] that enables the NaaVRE to 1) authenticate users and manages authentications when the users need to access remote infrastructures (e.g., cloud or HPC), and to 2) authorize users to access specific services and research assets of the VRE.

H. **The distributed data mesh** provides an interface for applications to share data content in remote infrastructure, e.g., via centralized storage or a peer-to-peer file system. It provides a transparent layer for workflow applications to exchange data.

I. **A distributed VRE ledger** that provides an immutable logging and ledger management framework for NaaVRE users to publish assets and trace the evolution of digital objects. We assume user communities are organized as a consortium of institutions or individuals. The VRE ledger is thus based on a consortium blockchain environment.

J. **Provenance and logs explorer** that allows the users to navigate the provenance information and system logs, analyze the specific events, and reproduce a workflow when necessary.

K. **The NaaVRE dashboard** will be the interface to configure the components, visualize the status of VRE components, e.g., the progress of the workflow execution, and utilization of the remote infrastructures.

L. **eScience DevOps** (development and operation) provides an automated pipeline to handle the lifecycle of the scientific code from code in the notebook cell to a self-contained and independent cloud computational entity.

Those components aim to support scientific activities identified in Fig. 1.

Figure 3: How NaaVRE supports scientific activities. The data management and infrastructure utilization parts are the same as Fig. 1.

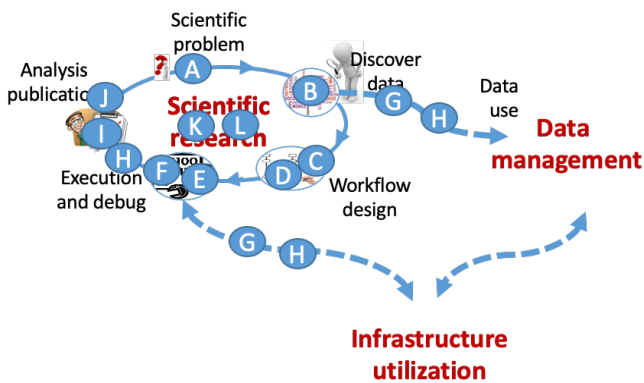

As shown in Fig. 3, a domain researcher can use NaaVRE knowledge base (A) to check existing solutions or similar problems in the community and search (B) existing data, software, and other relevant research assets for tackling the problem. The user can work on the experiment logic using the Jupyter environment or load existing code. The user can create reusable components by containerizing them as dockers (C), assembling them as workflow (D), and configuring the input data to create an experiment. The experiment will be executed by the workflow bus (E) in several steps, including planning infrastructure based on the workflow, automating the infrastructure services, and deploying the workflow component (F). In this context, the NaaVRE will automate the access to different infrastructures (including data repositories and cloud) using (G). The NaaVRE will also customize the data mesh for provisioning data for the execution (H). The runtime processes will be logged, and key processes will be tracked and recorded via the distributed ledger system (I). The user can analyze the experiment results using provenance and data analytics tools (J). The entire processes of the research activities will be presented via a dashboard (K) and automated using an eScience DevOps pipeline where needed (L).

The NaaVRE framework relies on external catalogs to access research assists from specific communities, e.g., the LifeWatch catalog for the ecology community, and the ENVRI-Hub catalog for the ENVRI community. The NaaVRE also relies on the external infrastructures to execute the remote workflows, including e-Infrastructure, e.g., EGI and EOSC, and cloud environments, e.g., Azure and AWS.

### C. How can NaaVRE be operated?

The NaaVRE can be seen both as a framework of the VRE technologies and as a community to enable users to collaborate via the Jupyter environment (Fig. 4).

Figure 4. The distributed research community is enabled by NaaVRE.

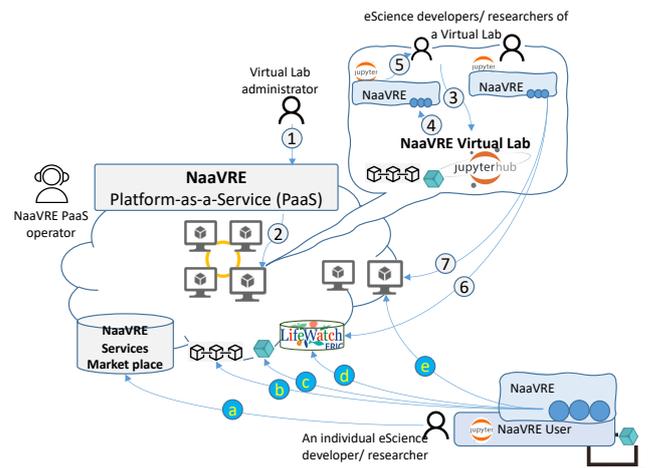

The NaaVRE can be operated by an organization as a Platform-as-a-Service (PaaS). Through the PaaS:

1. One (e.g., a Virtual Lab administrator) can create dedicated virtual labs (step 1) for a group of users, e.g., based on specific domain problems.
2. The NaaVRE PaaS will instantiate a Virtual Lab (VL) based on required resource capacity and configurations of the NaaVRE components, and deploy the VL in the infrastructure (step 2). A VL is typically based on the Jupyter Hub environment, with NaaVRE components installed.
3. VL users, e.g., eScience developers or researchers, will create their Jupyter environment via the Virtual Lab (step 3 and 4) and perform the research activities (step5)
4. Via the NaaVRE functional components, VL users can get access to the customized catalog (step 6) create dynamic resources (step 7) for running workflows.

5. The VL will initialize a blockchain node and P2P file node needed by distributed ledger and data mesh.

An individual user can also customize their Jupyter environment as private VL using NaaVRE technology (see the bottom part of Fig. 4)

1. A NaaVRE user can select relevant services from the NaaVRE marketplace and install them in the local environment (step a).
2. The users will also register themselves in the NaaVRE distributed ledger system (i.e., Blockchain) to access the ledgers (step b).
3. The user will install a Peer 2 Peer file node on the local environment to connect with the NaaVRE distributed data mesh. The shared layer can also be used for sharing data objects at the runtime of the workflow (step c).
4. Using the NaaVRE components, the user can then search assets (data or services) from the external catalog (e.g., from LifeWatch or ENVRI) to design an experiment (step d).
5. The code prototyped by the user in Jupyter will be executed on remote infrastructure via the distributed workflow bus by the experiment manager (step e).

## IV. IMPLEMENTATION AND CURRENT STATUS

The development of the NaaVRE follows the current industrial and community standards in cloud computing and scientific workflow management to develop the functional components.

### A. Development methodology

Some of the components are joint with the other projects, including SWITCH[11], VRE4EIC[12], ARTICONF[13], ENVRI-FAIR[14], and BlueCloud[15]. The development also follows the iterative and agile development software engineering practice, users are engaged in the development with different case studies.

The technical development team closely works with the domain scientists (the last three co-authors): discuss the use cases, identify user stories, define implementation plan, select technologies, test implementation via use cases and continue.

### B. Current prototype

The NaaVRE is developed as an open-source project[16]. Currently, the following components have been developed:

A. **VRE Knowledge base** is developed jointly with the ENVRI-FAIR project [10]. It is operated as a community service for the users of NaaVRE. The documents and metadata of the research assets will be indexed and captured in the knowledge base.
B. **VRE search engine.** The search engine is developed using information retrieval techniques [22] and provides an interface for VRE users to search assets (e.g., data sets, Web API or notebooks) from the community, and select relevant ones in a *basket*. The search engine provides an API for the Jupyter users to retrieve research assets from the *basket* in the cells of a notebook (Fig. 5).

Figure 5. the search engine prototyped using Elastic Search[17].

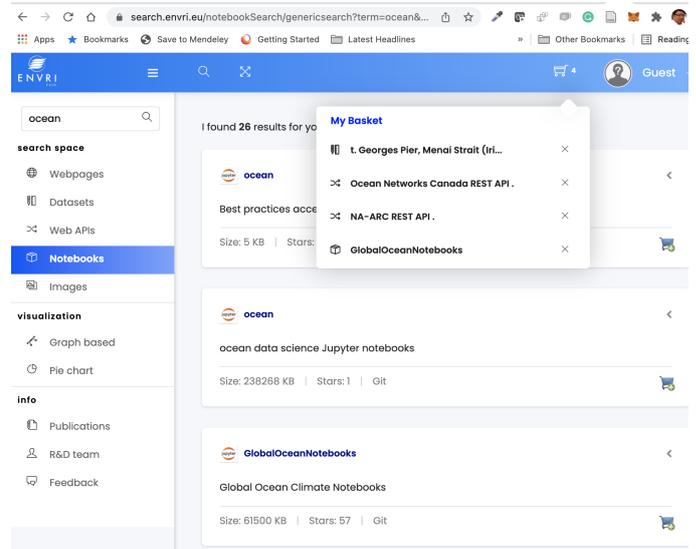

C. **Component Containerizer**, also known as FAIR-Cells extension [7], is an interactive tool for a researcher to 1) interactively encapsulate selected Cells of a notebook as RESTful services based on input and output variables of the Cells, 2) dockerize the services with customized base images (via Docker file), 3) publish the products (Docker images) in either a centralized way, namely on a remote repository (currently Docker Hub) and a catalog (e.g., LifeWatch catalog), or a decentralized way, namely using Blockchain (for metadata) and IPFS (for dock images). In this way, the notebook's fragments can reach better findability and accessibility towards the FAIR digital objects. Fig. 6 shows the screen snapshot of the FAIR-Cells extension. On the right side, the user can edit and select the relevant code fragments, and the tool will automatically capture the input and output of the Cell and visualize the component on the left side. From the left side panel, the user can generate the RESTful representation of the code in the Cell and dockerize it.

Figure 6. Screen snapshot of the component containerizer extension.

---

[11] http://switch-project.eu/
[12] https://vre4eic.ercim.eu/
[13] https://articonf.eu/
[14] http://envri.eu/envri-fair/
[15] https://www.blue-cloud.org/
[16] https://github.com/QCDIS/NaaVRE
[17] https://opensemanticsearch.org/

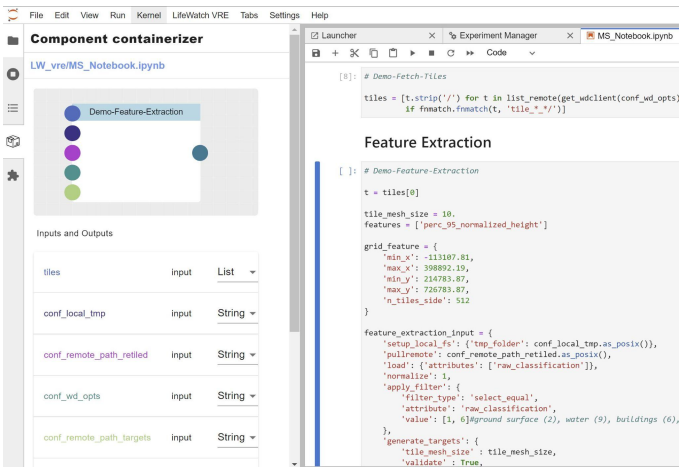

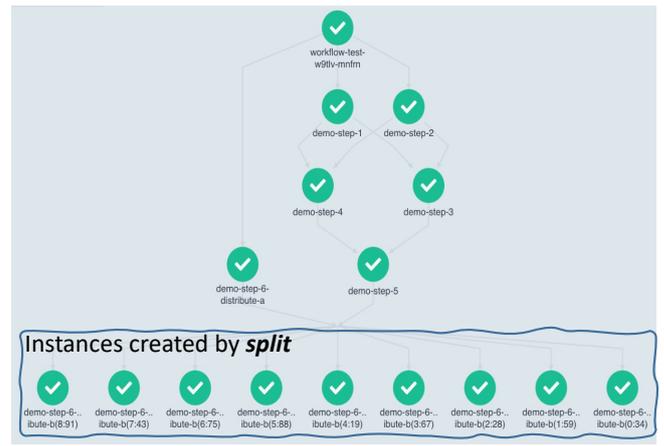

D. **Experiment manager**. The NaaVRE provides an effective way to allow users to describe the logic of the experiments (Fig. 7). The experiments can be based on the components created using the Containerizer tool. We use Common Workflow Language syntax to describe the workflow logic, which can be mapped onto different flow languages executed by a specific engine, e.g., ARGO flow or TOSCA. Using the Experiment Manager, a user can use two built-in components "split" and "merge" to parallelize the execution of certain cells on large data collection, as shown in Fig 7.

Figure 7. Screen snapshot of the workflow composer in the manager component.

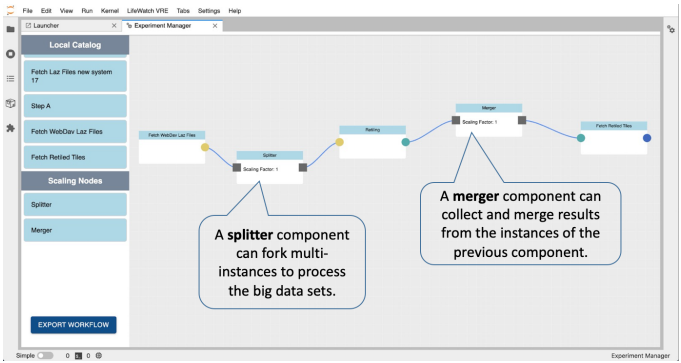

E. **Distributed workflow bus** is currently prototyped using ARGO framework. It interacts with the infrastructure automator to 1) prepare the cloud services and 2) automate the deployment and execution of an experiment. It also registers metadata and runtime information of the experiments on the distributed ledgers and enables collaborations among VRE users. Fig. 8 shows a screen snapshot of the experiment information in the current workflow bus. When the workflow contains "split" and "merge" operations, the workflow bus will dynamically create instances of the required workflow component based on provided data sources, and collect the output to a given data destination (in Fig. 8).

Figure 8. The runtime information of the distributed workflow bus.

F. **Remote infrastructure automator,** also called the Cloud-Cells extension, automates Cloud infrastructure provisioning and the deployment of Docker images on the Cloud infrastructure. Cloud-Cells uses SDIA (the Software Defined Infrastructure Automator), a framework developed by the same team for planning, provisioning, and deploying applications on multi-Cloud environments [11] (Fig. 9). A VRE can have a number of pre-provisioned virtual infrastructure (namely networked virtual machines); it can dynamically request and provision virtual infrastructure requested by a user for running specific workflows. A user should provide his/he own credential to obtain the dynamic resources. A dynamic infrastructure can be released after the workflow is finished.

Figure 9. The screen snapshot of the Infrastructure Automator and Experiment Manager.

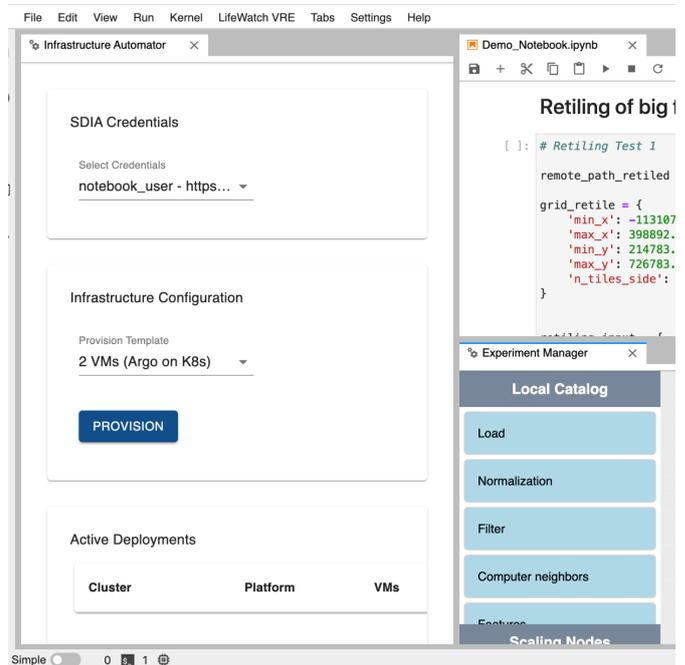

G. **Authentication and authorization infrastructure (AAI)** currently follows the best practices and guidelines from the initiatives like EOSC, ENVRI, and LifeWatch. It currently supports the public identity providers like GitHub. Virtual

Lab administrators can log in to the NaaVRE platform using the identities.

H. **The distributed data mesh** currently supports both centralized (WebDAV[18]) and decentralized (IPFS[19]) paradigms.

I. **The distributed VRE ledger** currently uses HyperLedger Fabric[20]. Instead of using a public blockchain, we design the ledger of a community as a consortium of a set of trusted organizations. A new organization of the NaaVRE can set up a new node and join the network of the VRE ledger. Individual users can join the network using the identity of their organization. Other identity providers will also be supported in the future.

J. **Provenance explorer and analysis**. The tool provides an interface for users to interactively explore the system logs and the provenance of the workflow and to identify the anomalies and reproduce the workflows or problems [12]. The provenance of the workflow is captured using PROV-O[21]. Via the explorer, researchers will be able to interactively explore the provenance (see the upper part of Fig. 10) of the workflow among cells (created using Experiment manager) together with the workflow processes and system logs (see the middle and bottom part of Fig. 10), and identify the anomalies during the execution.

Figure 10. The provenance explorer and analysis interface. A user can visualize the provenance graph, workflow process and infrastructure performance logs.

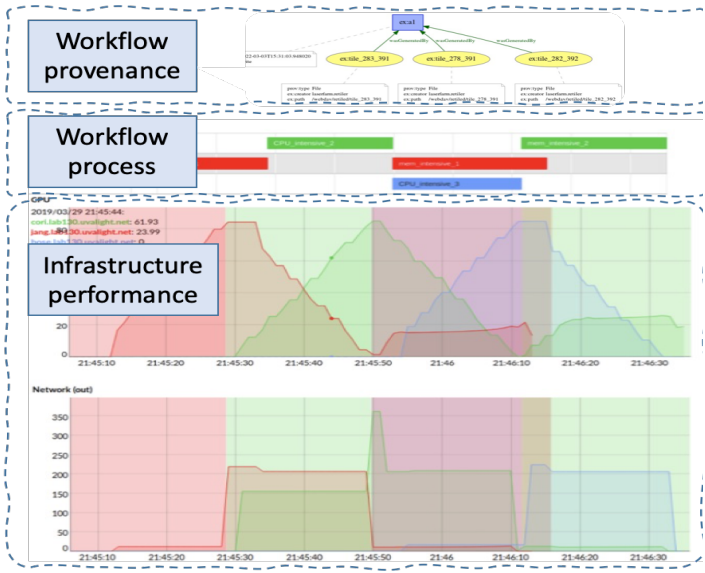

K. **The NaaVRE dashboard** visualizes the runtime status of the scientific experiments, infrastructure status, and distributed ledgers; currently, the ARGO framework is used to prototype the dashboard.

L. **eScience DevOps** applies DevOps practices, including continuous integration (CI) and continuous deployment (CD) to automate the lifecycle of the experiments in the VRE. The NaaVRE uses Git to manage versions of the source of the cell, and DockerHub or IPFS to store the images of containerized cells. The containerization and orchestration of the cells in a workflow on remote cloud infrastructure can be automated via the eScience DevOps pipeline when new changes are made on the source.

V. CASE STUDIES: *LidarAirCloud virtual lab*[22]

Airborne Laser Scanning (ALS) data derived from Light Detection and Ranging (LiDAR) technology allow the construction of Essential Biodiversity Variables (EBVs) of ecosystem structure with high resolution at the landscape, national and regional scales [14]. Researchers nowadays often process LiDAR data for various ecological applications, and rapidly prototype using script languages like R or python and share their experiments via scripts or more recently via notebook environments, such as Jupyter.

A legacy program called 'Laserchicken'[13] was developed in a previous project to process country-wide LiDAR point clouds in a local environment (e.g., the Dutch national ICT infrastructure called SURF). It is a Python tool that focuses on extracting features (i.e. statistical properties) of user-defined subsets of point cloud data [13]. The basic workflow of Laserchicken is shown in Fig. 11. It consists of four core modules (load, compute neighbors, features, export) (as shown in Fig. 11 in blue color), and two optional modules (filter, normalize) (as shown in Fig. 11 in grey color), which provide a processing workflow for feature extraction from LiDAR point clouds, with file-based input and output.

Fig. 11 Laserchicken workflow for processing LiDAR point clouds

---

[18] http://www.webdav.org/
[19] https://ipfs.io/
[20] https://www.hyperledger.org/use/fabric
[21] https://www.w3.org/TR/prov-o/
[22] The development of the use case is partially supported by the EOSC ENVRI-FAIR early adopter program (EAP) project, and the Azure AI on earth project.

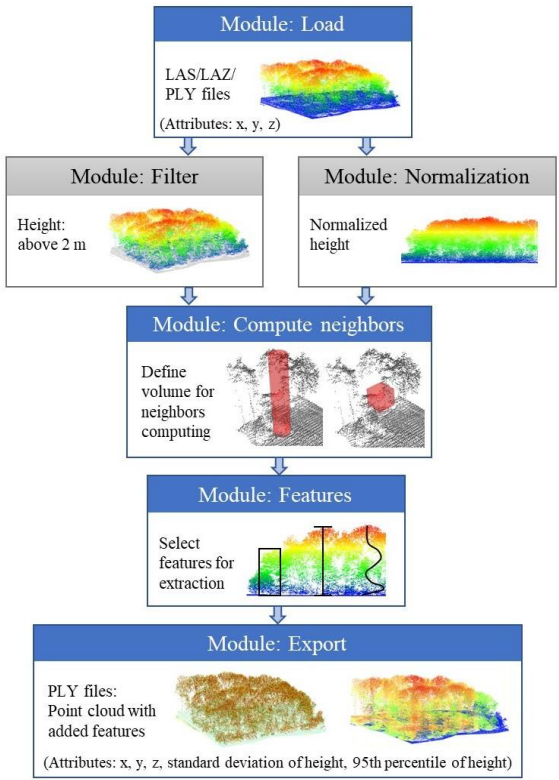

However, a user currently can only use the Laserchicken application either on the dedicated infrastructure or on the local machine to process the LiDAR data. Running a specific part of the code on remote infrastructure is often challenging for domain scientists. The capacity of the local machine or a given infrastructure also limits the volume of data. This use case (LidarAirCloud) is to demonstrate how the NaaVRE solution enables scientists to scale experiments to large data volumes, different data sources, or new models.

Using the NaaVRE, the user first needs to set up the environment, e.g., selecting the following components: *Experiment manager, component containerizer, distributed workflow bus, remote infrastructure automator, distributed data mesh, and distributed ledger*, as shown in the upper part of Fig. 11. Since not all the components have been integrated, only a small subset has been used.

Figure 12. The workflow of the use case.

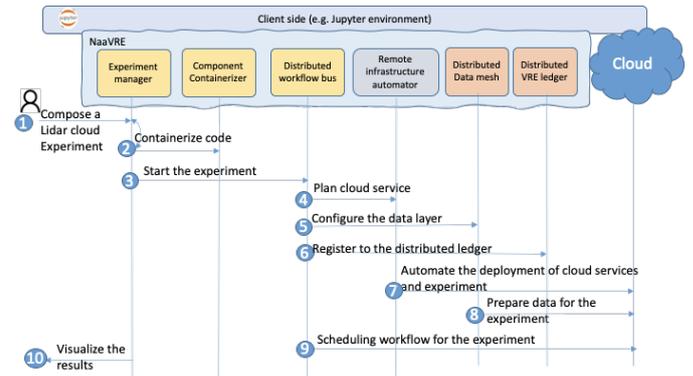

In this simple example, the user does not have to search for data from an extensive collection of the data sources; data sets from Table 1 have been considered.

| Country | Multi-year | Point density | Data volume* |
|---|---|---|---|
| Finland | Partly | 1–2 pt/m$^2$ | 4 TB |
| Netherlands | Yes | 0.1–20 pt/m$^2$ | 16 TB |
| Spain | Partly | 0.5–2 pt/m$^2$ | 5 TB |
| Denmark | Yes | 4–5 pt/m$^2$ | 8 TB |
| Estonia | Yes | 0.2–18 pt/m$^2$ | - |
| Slovenia | Yes | 2-5pt/m$^2$ | 2.5 TB |
| UK | Partly | 0.5-16 pt/m$^2$ | 45 TB |

Table 1: Details of LiDAR data from example countries in Europe. Data volume represents how much data storage is needed, based on the number of files available in each download portal and the average size of each file. The data used in this paper is from the Netherland[23].

The sequences under the NaaVRE components in Fig. 11 demonstrate the basic steps in the use case.

1. After preparing the LiDAR data that need to be processed, the user can use **Experiment manager** to design an experiment based on the 'Laserchicken' workflow in the Jupyter environment. In this use case, the workflow was designed as the following steps: first, loading the raw LiDAR data (LAS files) covering a sample area in the Netherlands; second, normalizing the height of the point cloud; third, computing neighbors with a target volume of an infinite cylinder at 1-meter radius; fourth, extracting selected features (e.g., 95th percentile of normalized height, coefficient of variance of normalized height); and finally, exporting the enhanced point cloud with added features (PLY files).

2. Via the experiment manager, the user will need to containerize the code as portable containers using the **containerizer** component provided in the Jupyter environment. Each step of the user's workflow described above can be containerized (and then published) as a container, providing a specific function of processing

---

[23] https://www.ahn.nl/ahn-viewer

LiDAR data. This and the previous step can be interactive until a concrete experiment is composed. An experiment can be seen as a concrete workflow with configurations on input data and other necessary parameters.
3. After the experiment is ready, the user will submit the experiment to **the distributed workflow bus** for execution.
4. **The distributed workflow bus** will then plan cloud services (e.g., Virtual Machines and their topology) for running the workflow by using the **Remote infrastructure automator**.
5. The **distributed workflow bus** will configure the **distributed data mesh** needed by the experiment, e.g., centralized (based on WebDav) or decentralized (based on IPFS) option.
6. The **distributed workflow bus** will also register the experiment in the **VRE ledger.**
7. The **remote infrastructure automator** will automate the provision of the cloud services on selected providers. At this stage, the component of AAI will be used.
8. The **distributed data mesh** will also provision the data sets needed by the workflow. In this use case, pre-designed workflow can be upscaled and applied to other country-wide LiDAR datasets (with multi-terabyte data volume), as shown in Table 1. Ecosystem structure products covering whole countries or regions can be generated and used for further analysis.
9. The **distributed workflow bus** will schedule the workflow execution after components are being deployed (in step 4) and data is being prepared (in step 6).
10. After the execution, the user can visualize the results via the notebook. Fig. 13 shows an example visualization of generated data products of the Netherlands from the Laserchicken workflow.

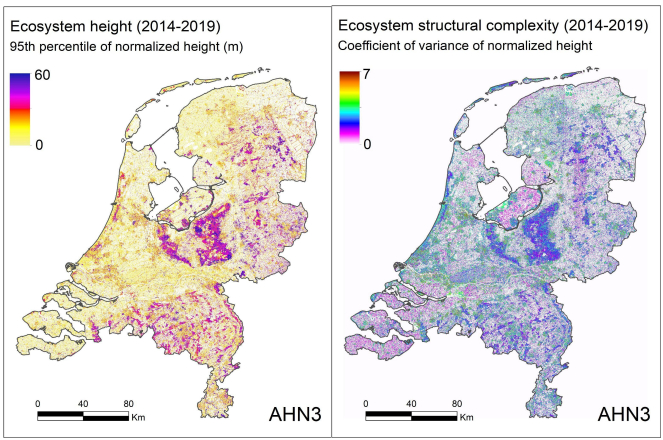

Figure 13. The visualization of two data products of ecosystem structure (height & structural complexity) of the Netherlands from the Laserchicken workflow. The LiDAR data were collected during the third Dutch national airborne laser scanning flight campaign (AHN3, *Actueel Hoogtebestand Nederland*).

The initial version of the NaaVRE components has been tested on the cloud environment (offered by the EOSC early adopter program for ENVRI-FAIR, and Microsoft Azure through the AI for Earth program) to process a much bigger dataset than in a local environment [16].

## VI. DISCUSSION

In this use case, we demonstrate how a user can use NaaVRE to enhance the Jupyter lab to execute the legacy Laserchicken workflow in remote infrastructures. In the paper, the target users are mainly developers or researchers using eScience approaches. They normally have interdisciplinary knowledge across scientific domains and computing techniques. They use the Jupyter environment for developing scientific code; the eScience researchers focus more on the scientific output, while the eScience developers are dedicated programmers for scientific code and often act as support staff for eScience or regular domain researchers.

The NaaVRE architecture aims to be open and flexible for other systems on workflow management. From the current prototype, we have specifically considered the following aspects.

*Connect with other workflow management systems.* NaaVRE aims to support other workflow systems besides the default one included in the marketplace as an open ecosystem. The NaaVRE can execute the workflows supported by an external workflow management system via the workflow bus, where the description will be wrapped and executed by invoking the external engine. The distributed workflow bus can dynamically launch the external engine when no such running engine instance is accessible. However, this integration also requires the external workflow management system to provide a backend engine with a well-defined interface for remote invocations.

On the other hand, *the distributed workflow bus and the remote infrastructure automator are decoupled* from the user interface of the Notebook. In principle, the backend can be integrated with the workflow composition interface of the third-party workflow management interface for designing and executing the workflow. However, this integration requires efforts for adapting the external composition interface to generate the required workflow description of the NaaVRE and to invoke the workflow bus and Remote infrastructure automator in NaaVRE.

The above two options give opportunities for applying NaaVRE in the current service architecture in the LifeWatch ERIC, besides operating the NaaVRE as an online Platform-as-a-Service.

## VII. SUMMARY

The NaaVRE framework is jointly developed in the context of ENVRI-FAIR, LifeWatch, Clarify, and several other projects. We closely interact with the users from ecology, marine, and other domains in the environmental and earth sciences. The framework is developed based on the Jupyter environment. The architecture aims to deliver an open ecosystem for the VRE components needed by the users. The framework is also going to be used as a framework for supporting distributed machine learning workflows in the medical domain[24].

ACKNOWLEDGMENT

This work has been partially funded by the European Union's Horizon 2020 research and innovation program by the ENVRI-FAIR project grant agreement No 824068, by the BLUECLOUD project grant agreement No 862409, by the LifeWatch ERIC, by the ARTICONF project grant agreement No 825134, and by the project CLARIFY under the Marie Sklodowska-Curie grant agreement No 860627. The use case has also been supported by an AI for Earth Grant (AI4E-1111-Q3N7-20100806) from Microsoft which provided access to the Azure Cloud in the context of the LidarAirCloud project. The Laserchicken workflow was developed in the eEcoLiDAR project funded by the Netherlands eScience Center (grant number ASDI.2016.014).

---

[24] http://www.clarify-project.eu